\documentstyle[prl,aps,floats,epsf]{revtex}
\newcommand{\met}{$\rlap{\kern0.25em/}E_T$}
\newcommand{\gmet}{$\gamma\rlap{\kern0.25em/}E_T$}
\newcommand{\gmetjj}{$\gamma\rlap{\kern0.25em/}E_T\ +\ge 2\ {\rm jets}$}
\newcommand{\gmetnj}{$\gamma\rlap{\kern0.25em/}E_T+n\ {\rm jets}$}
\newcommand{\gjj}{$\gamma\ +\ge 2\ {\rm jets}$}
\newcommand{\emetjj}{$e\rlap{\kern0.25em/}E_T\ +\ge 2\ {\rm jets}$}
\begin{document}

%
\author{                                                                      
B.~Abbott,$^{40}$                                                             
M.~Abolins,$^{37}$                                                            
V.~Abramov,$^{15}$                                                            
B.S.~Acharya,$^{8}$                                                           
I.~Adam,$^{39}$                                                               
D.L.~Adams,$^{48}$                                                            
M.~Adams,$^{24}$                                                              
S.~Ahn,$^{23}$                                                                
H.~Aihara,$^{17}$                                                             
G.A.~Alves,$^{2}$
% ...
\\                                                                            
\vskip 0.70cm                                                                 
\centerline{(D\O\ Collaboration)}                                             
\vskip 0.70cm                                                                 
}                                                                             
\address{                                                                     
\centerline{$^{1}$Universidad de Buenos Aires, Buenos Aires, Argentina}       
\centerline{$^{2}$LAFEX, Centro Brasileiro de Pesquisas F{\'\i}sicas,         
                  Rio de Janeiro, Brazil}                                     
\centerline{$^{3}$Universidade do Estado do Rio de Janeiro,                   
                  Rio de Janeiro, Brazil}                                     
\centerline{$^{4}$Universidad de los Andes, Bogot\'{a}, Colombia}             
\centerline{$^{5}$DAPNIA/Service de Physique des Particules, CEA, Saclay,     
                  France}                                                     
\centerline{$^{6}$Panjab University, Chandigarh, India}                       
\centerline{$^{7}$Delhi University, Delhi, India}                             
\centerline{$^{8}$Tata Institute of Fundamental Research, Mumbai, India}      
\centerline{$^{9}$Kyungsung University, Pusan, Korea}                         
\centerline{$^{10}$Seoul National University, Seoul, Korea}                   
\centerline{$^{11}$CINVESTAV, Mexico City, Mexico}                            
\centerline{$^{12}$Institute of Nuclear Physics, Krak\'ow, Poland}            
\centerline{$^{13}$Institute for Theoretical and Experimental Physics,        
                   Moscow, Russia}                                            
\centerline{$^{14}$Moscow State University, Moscow, Russia}                   
\centerline{$^{15}$Institute for High Energy Physics, Protvino, Russia}       
\centerline{$^{16}$University of Arizona, Tucson, Arizona 85721}              
\centerline{$^{17}$Lawrence Berkeley National Laboratory and University of    
                   California, Berkeley, California 94720}                    
\centerline{$^{18}$University of California, Davis, California 95616}         
\centerline{$^{19}$University of California, Irvine, California 92697}        
\centerline{$^{20}$University of California, Riverside, California 92521}     
\centerline{$^{21}$Florida State University, Tallahassee, Florida 32306}      
\centerline{$^{22}$University of Hawaii, Honolulu, Hawaii 96822}              
\centerline{$^{23}$Fermi National Accelerator Laboratory, Batavia,            
                   Illinois 60510}                                            
\centerline{$^{24}$University of Illinois at Chicago, Chicago,                
                   Illinois 60607}                                            
\centerline{$^{25}$Northern Illinois University, DeKalb, Illinois 60115}      
\centerline{$^{26}$Northwestern University, Evanston, Illinois 60208}         
\centerline{$^{27}$Indiana University, Bloomington, Indiana 47405}            
\centerline{$^{28}$University of Notre Dame, Notre Dame, Indiana 46556}       
\centerline{$^{29}$Purdue University, West Lafayette, Indiana 47907}          
\centerline{$^{30}$Iowa State University, Ames, Iowa 50011}                   
\centerline{$^{31}$University of Kansas, Lawrence, Kansas 66045}              
\centerline{$^{32}$Louisiana Tech University, Ruston, Louisiana 71272}        
\centerline{$^{33}$University of Maryland, College Park, Maryland 20742}      
\centerline{$^{34}$Boston University, Boston, Massachusetts 02215}            
\centerline{$^{35}$Northeastern University, Boston, Massachusetts 02115}      
\centerline{$^{36}$University of Michigan, Ann Arbor, Michigan 48109}         
\centerline{$^{37}$Michigan State University, East Lansing, Michigan 48824}   
\centerline{$^{38}$University of Nebraska, Lincoln, Nebraska 68588}           
\centerline{$^{39}$Columbia University, New York, New York 10027}             
\centerline{$^{40}$New York University, New York, New York 10003}             
\centerline{$^{41}$University of Rochester, Rochester, New York 14627}        
\centerline{$^{42}$State University of New York, Stony Brook,                 
                   New York 11794}                                            
\centerline{$^{43}$Brookhaven National Laboratory, Upton, New York 11973}     
\centerline{$^{44}$University of Oklahoma, Norman, Oklahoma 73019}            
\centerline{$^{45}$Brown University, Providence, Rhode Island 02912}          
\centerline{$^{46}$University of Texas, Arlington, Texas 76019}               
\centerline{$^{47}$Texas A\&M University, College Station, Texas 77843}       
\centerline{$^{48}$Rice University, Houston, Texas 77005}                     
}                                                                             

\date{August 13, 1998}

\title{Search for Squarks and Gluinos in Single-Photon Events with Jets and 
       Large Missing Transverse Energy in $p\bar{p}$ Collisions at 
       $\sqrt{s}=1.8$~TeV}

\maketitle

\begin{abstract}
  We search for physics beyond the standard model using events with a photon, 
  two or more hadronic jets, and an apparent imbalance in transverse energy,
  in $p\bar{p}$ collisions at the Fermilab Tevatron at $\sqrt{s}=1.8$~TeV. 
  Such events are predicted for production of supersymmetric particles. 
  No excess is observed beyond expected background. For the parameter space 
  of the minimal supersymmetric standard model with branching fraction 
  ${\rm B}(\tilde\chi^0_2\rightarrow\gamma\tilde\chi^0_1)=1$ and 
  $m_{\tilde\chi^0_2}-m_{\tilde\chi^0_1}>20$~GeV, we obtain a 95\% 
  confidence level lower limit of 310 GeV for the masses of squarks and 
  gluinos, where their masses are assumed equal.
\end{abstract}


\twocolumn
We search for physics beyond the standard model~(SM) using events with one high 
transverse energy ($E_T$) photon, two or more jets, and large imbalance in 
transverse energy (\met). We call these \gmetjj\ events. This search is 
motivated by recent suggestions\cite{gmsb,kane} that supersymmetry may result 
in signatures involving one or more photons together with multiple jets and 
large \met. 

Supersymmetry is a generalization of space-time symmetry. It introduces for 
every particle in the standard model a supersymmetric partner differing in spin 
by one half. {\it R}-parity\cite{rpar}, defined as $+1$ for SM particles and 
$-1$ for their super-partners, is assumed to be conserved in this analysis,
such that supersymmetric particles are produced in pairs and
the lightest supersymmetric particle~(LSP) is stable.
In the minimal supersymmetric standard model (MSSM), the gaugino-Higgsino
sector (excluding gluinos) is described by four parameters: $M_1$, $M_2$,
$\mu$, and $\tan\beta$, where $M_1$ and $M_2$ are the U(1) and 
SU(2) gaugino mass parameters, $\mu$ is the Higgsino mass parameter,
and $\tan\beta$ is the ratio of the vacuum expectation values of the 
two Higgs doublets. Gaugino-Higgsino mixing gives  
four neutral mass eigenstates (neutralinos $\tilde\chi^0_i,\ i=1,...,4$) 
and two charged mass eigenstates (charginos $\tilde\chi^\pm_i,\ i=1,2$).
Within the MSSM, the radiative decay of 
$\tilde\chi^0_2\rightarrow\gamma\tilde\chi^0_1$ dominates in the region 
$50\stackrel{<}{_\sim}M_1\sim M_2\stackrel{<}{_\sim}100\ {\rm GeV}$, 
$1\stackrel{<}{_\sim}\tan\beta\stackrel{<}{_\sim}3$ and 
$-65\stackrel{<}{_\sim}\mu\stackrel{<}{_\sim}-35\ {\rm GeV}$
of parameter space\cite{pspace}, and has been proposed as an
explanation\cite{kane} of a candidate event reported by the CDF 
Collaboration\cite{cdf}. Assuming that $\tilde\chi^0_1$ is the LSP, 
then the production of $\tilde\chi^0_2$, either directly or from 
decays of other supersymmetric particles, will yield 
$\gamma\rlap{\kern0.25em/}E_T + X$ events. 

In this Letter, we present a search for physics beyond the SM in the channel 
$p\bar{p}\rightarrow\gamma\rlap{\kern0.25em/}E_T\ +\ge 2\ {\rm jets}$ 
at the Fermilab Tevatron collider. Because of large backgrounds from QCD
processes, we do not consider events with less than two jets.
We interpret our results in terms of squark ($\tilde q$)
and gluino ($\tilde g$) production in the context of supersymmetric models 
with a dominant $\tilde\chi^0_2\rightarrow\gamma\tilde\chi^0_1$ decay.


The data used in this analysis were collected with the D\O\ detector during the
1992--1996 Tevatron run at a center of mass energy of $\sqrt{s}$ = 1.8~TeV, 
and represent an integrated 
luminosity of $99.4\pm 5.4$~pb$^{-1}$. A detailed description of the D\O\
detector can be found in Ref.\cite{dzero}. The trigger requires one 
electromagnetic (EM) cluster with $E_T>15$~GeV, one jet 
with $E_T>10$~GeV, and $\rlap{\kern0.25em/}E_T > 14$~GeV 
($\rlap{\kern0.25em/}E_T > 10$~GeV for about 10\% of the data taken early 
in the Tevatron run). 
Photons are identified via a two-step process: the selection of isolated 
EM energy clusters, and the rejection of such clusters with any associated 
charged tracks. The EM clusters are 
selected from calorimeter energy clusters by requiring:
(i)   at least 95\% of the energy to be deposited in the EM section
      of the calorimeter;
(ii)  the transverse and longitudinal shower profiles to be consistent
      with those expected for an EM shower; and
(iii) the energy in an annular isolation cone with radius 
      ( ${\cal R}\equiv\sqrt{(\Delta\phi)^2+(\Delta\eta)^2}$ ) 
      0.2 to 0.4 around the cluster in $\eta-\phi$ space to be less than 
      10\% of the EM energy in an ${\cal R}=0.2$ cone, where $\eta$ and $\phi$ 
      are the pseudorapidity and azimuth, respectively.
The EM clusters that have either a reconstructed track or a large number 
of hits in the tracking chamber along a road joining the cluster and the 
interaction vertex are vetoed. $\rlap{\kern0.25em/}E_T$~is determined
from the energy deposition in the calorimeter within $|\eta|<4.5$.

\begin{figure}[htbp]
  \epsfysize=3.0in\epsfbox{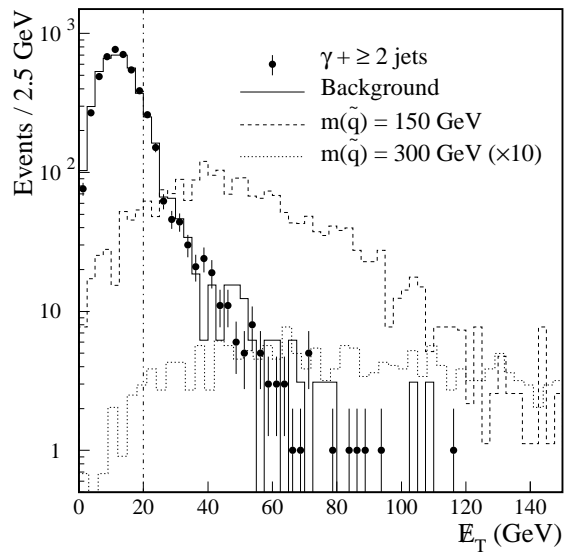}
  \caption{The \protect\met\ distributions of the \protect\gjj\ (solid circles)
           and background (solid histogram) events.
           The number of events in the background is normalized to the 
           \protect\gjj\ sample for $\rlap{\kern0.25em/}E_T < 20$~GeV, the
           region left of the dot-dashed line.
           Also shown (dashed and dotted histograms) are the distributions 
           expected from supersymmetry for 
           $m_{\tilde q}=m_{\tilde g}=150$~GeV and 300 GeV.}
    \label{fig:prl1}
\end{figure}

To be selected as \gmetjj\ candidates, events are first required 
to have at least one identified photon with $E_T^\gamma>20$~GeV and 
pseudorapidity $|\eta^\gamma|<1.1$ or $1.5<|\eta^\gamma|<2.0$, and two 
or more jets reconstructed with cones of radius ${\cal R}=0.5$, having
$E^j_T>20$~GeV and $|\eta^j|<2.0$. We refer to the events passing these 
requirements as the \gjj\ sample. The \met\ distribution of these events 
is shown in Fig.~\ref{fig:prl1}. We then require 
$\rlap{\kern0.25em/}E_T>25$~GeV. A total of 318 events satisfy all 
requirements. 

%
%

The principal backgrounds to the signal are: events from sources such as QCD 
direct photon and multijet events, where there is mismeasured \met\ and
a real or fake photon;
$W(\rightarrow e\nu)+{\rm jets}$ events, 
where the electron is misidentified as a photon; 
and $W(\rightarrow\ell\nu)+{\rm jets}$ events (where $\ell=e,\mu,\tau$), 
in which one of the jets is misidentified as a photon. These backgrounds
are estimated from the data sample with the same trigger as the candidate
events. The background from mismeasurement of $\rlap{\kern0.25em/}E_T$\ 
is estimated using events with one EM-like cluster that satisfies all 
photon criteria, except requirement (ii) on the shower profile.
These events must also have two or more jets with $E^j_T>20$~GeV 
and $|\eta^j|<2.0$, making them similar to those of the \gjj\ sample, and
therefore of similar resolution in $\rlap{\kern0.25em/}E_T$. The events 
in this background sample are normalized to the \gjj\ sample for
$\rlap{\kern0.25em/}E_T < 20$~GeV, which provides an estimated background 
from \met\ mismeasurement of $315\pm 30$ events beyond 
$\rlap{\kern0.25em/}E_T = 25$~GeV.

$W\ +\ge 2\ {\rm jets}$ events with $W\rightarrow e\nu$ can mimic \gmetjj\ 
events if the electron is misidentified as a photon. This contribution
is estimated using a sample of \emetjj\ events that passes all our kinematic 
requirements, with the electron satisfying those defined for the photon.
Electrons are selected from identified EM clusters that have matched tracks.
The probability that an electron is misidentified as a photon is determined 
from $Z\rightarrow ee$ events as $0.0045\pm 0.0008$. Multiplying this 
probability by the number of \emetjj\ events yields a background of $4\pm 1$ 
events.

The $W(\rightarrow\ell\nu)+{\rm jets}$ background is estimated
using a data sample of $W(\rightarrow e\nu)\ +\ge 3\ {\rm jets}$ events 
passing all kinematic requirements, with at least one of the jets 
satisfying those imposed on photons. The probability that a jet is 
misidentified as a photon is determined by counting the number of photons 
observed in multijet events. We find this to be $0.0007\pm 0.0002$. Using this 
probability and the scale factor $N_{W(\rightarrow\ell\nu)+\ge 3\ {\rm jets}}/
N_{W(\rightarrow e\nu)+\ge 3\ {\rm jets}}$ (determined from Monte Carlo), 
we estimate a background of $1.0\pm 0.3$ events. The background from 
$Z(\rightarrow\nu\nu)\ +\ge 3\ {\rm jets}$ is found to be negligible.

\begin{table}[htbp]
  \squeezetable
  \begin{tabular}{c|cc|cc|cc}
   Number &\multicolumn{2}{c|}{\met$ > 25$ GeV} 
          &\multicolumn{2}{c|}{\met$ > 25$ GeV} 
          &\multicolumn{2}{c}{\met$ > 50$ GeV} \\
   of     &\multicolumn{2}{c|}{No $H_T$ cut} 
          &\multicolumn{2}{c|}{$H_T>200$ GeV} 
          &\multicolumn{2}{c}{No $H_T$ cut} \\
  jets    & $N_S$ & $N_B$ & $N_S$ & $N_B$ & $N_S$ & $N_B$ \\ \hline
 $n\ge 2$ & 318 & 320$\pm$30 & 30 & 20$\pm$10 & 43 & 65$\pm$15 \\
 $n\ge 3$ &  70 &  70$\pm$15 & 17 &  8$\pm$5  & 11 & 10$\pm$5 \\
 $n\ge 4$ &   8 &  10$\pm$5  &  6 &  4$\pm$3  &  1 &  3$\pm$3 \\
  \end{tabular}
  \caption{Number of observed \protect\gmetnj\ events ($N_S$) together 
           with the corresponding number of background events ($N_B$) 
           for $n\ge 2, 3, 4$, for three sets of cutoffs.}
  \label{tab:data}
\end{table}

\begin{figure}[htbp]
  \epsfysize=3.0in\epsfbox{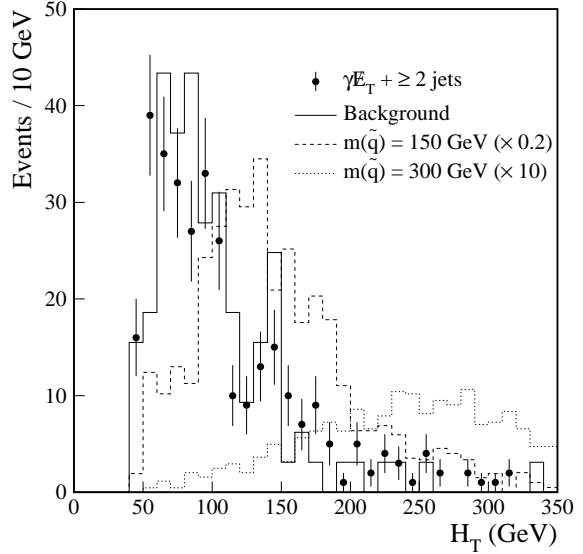} 
  \caption{The $H_T$ (defined as $\sum_j E^j_T$) distributions of the 
           \protect\gmetjj\ and background events. The expected distributions 
           from supersymmetry are also shown for comparison.}
  \label{fig:prl2}
\end{figure}

The number of observed events and the expected backgrounds are summarized in 
Table~\ref{tab:data}, together with breakdowns into events 
with three or more and four or more jets. The $H_T$ distribution (defined as 
the scalar sum of the $E_T$ of all jets with $E^j_T>20$~GeV and $|\eta^j|<2.0$) 
is shown in Fig.~\ref{fig:prl2}, for both \gmetjj\ and background 
samples. The background distribution is consistent with that observed for 
\gmetjj. Also given in Table~\ref{tab:data} is the number of observed events 
and the expected background if the cutoff $H_T>200$~GeV is applied or if 
the \met\ cutoff is raised to 50~GeV. In all three comparisons, 
the estimated number of background events agrees with the number of 
events observed in the data. 

To optimize selection criteria for a supersymmetric signal, we simulate 
squark and gluino pair production, and also production in association with 
charginos or neutralinos using the {\sc spythia} program\cite{pythia}. 
The MSSM parameters are set to 
$M_1=M_2=60$~GeV, $\tan\beta=2$, and $\mu=-40$~GeV.
This set gives $m_{\tilde\chi^0_1}=34$~GeV,
$m_{\tilde\chi^0_2}=60$~GeV, and
${\rm B}(\tilde\chi^0_2\rightarrow\gamma\tilde\chi^0_1)=1$.
Sleptons ($\tilde\ell$) and stop ($\tilde t_1$) are assumed to be heavy. 
Events with $\tilde\chi^0_2$ in the final state are selected and processed 
through the D\O\ detector-simulation program~\cite{geant}, and the trigger 
simulator. The same trigger requirements, reconstruction, and selection
criteria are then applied as were used with the data. 
Monte Carlo~(MC) events are generated for three squark or gluino mass 
possibilities: 
(i) equal mass squark and gluino ($m_{\tilde q}=m_{\tilde g}$);
(ii) heavy squark and light gluino ($m_{\tilde q}\gg m_{\tilde g}$); and
(iii) light squark and heavy gluino ($m_{\tilde q}\ll m_{\tilde g}$).
The \met\ and $H_T$ distributions for
$m_{\tilde q} = m_{\tilde g} = 150, 300$~GeV events are shown, respectively,
in Figs.~\ref{fig:prl1} and \ref{fig:prl2}, where the MC distributions are
scaled by the factors shown in parentheses.
The distributions expected from supersymmetry differ considerably from 
those of the background. To increase the sensitivity to supersymmetry, 
we introduce an $H_T$ cutoff, and maximize the 
$\epsilon_S/\delta N_B$ ratio by varying the \met\ and $H_T$ cutoffs. Here 
$\epsilon_S$ is the efficiency for signal, and $\delta N_B$ is the 
uncertainty on the estimated number of background events. 
To ensure high efficiencies for both low and high squark and gluino masses,
the optimization is done for two MC points 
$m_{\tilde q}=m_{\tilde g}=150\ {\rm and}\ 300$~GeV.
The optimum values are \met$ > 35$~GeV and $H_T>100$~GeV
for 150~GeV, and \met$ > 45$~GeV and $H_T>220$~GeV
for 300~GeV. The $\epsilon_S/\delta N_B$ results (a function of 
squark/gluino mass) for the two sets of optimized cutoffs are equal near 
200~GeV. Therefore we apply the cutoffs 
optimized for the 150~GeV mass point to MC events with squark and gluino 
masses below 200~GeV, and apply those optimized for the 300~GeV mass
point to masses of 200~GeV or above. The number of 
events observed for these two sets of cutoffs are 60 and 5, with 
$75\pm 17$ and $8\pm 6$ events expected from background processes. 
We consequently observe no excess beyond the standard model.

\begin{table}[htbp]
  \squeezetable
  \begin{tabular}{c|cc|cc|cc}
   $m_{{\tilde q}/{\tilde g}}$ &\multicolumn{2}{c|}{$m_{\tilde q}(=m_{\tilde g})$}
                               &\multicolumn{2}{c|}{$m_{\tilde g}(\ll m_{\tilde q})$}
                               &\multicolumn{2}{c} {$m_{\tilde q}(\ll m_{\tilde g})$} \\
   GeV & $\epsilon_0$ (\%) & $\epsilon_S$ (\%) & $\epsilon_0$ (\%) & $\epsilon_S$ (\%) 
       & $\epsilon_0$ (\%) & $\epsilon_S$ (\%) \\ \hline
    150 & 66.2 & 15.1$\pm$0.8 & 69.1 & 11.6$\pm$0.9 & 60.0 & 16.8$\pm$1.1 \\
    200 & 62.3 &  7.9$\pm$0.6 & 59.6 &  5.3$\pm$0.6 & 53.8 &  9.5$\pm$0.9 \\
    250 & 59.6 & 14.8$\pm$0.8 & 49.7 & 13.6$\pm$1.1 & 55.4 & 14.8$\pm$1.1 \\
    300 & 56.1 & 21.5$\pm$1.0 & 43.1 & 19.0$\pm$1.3 & 55.4 & 22.1$\pm$1.2 \\
    350 & 51.8 & 22.8$\pm$1.1 & 39.3 & 23.5$\pm$1.5 & 52.7 & 26.6$\pm$1.4 \\
    400 & 46.7 & 23.5$\pm$1.1 & 33.3 & 22.7$\pm$1.6 & 54.3 & 25.8$\pm$1.3 \\
  \end{tabular} 
  \caption{The percentages of events ($\epsilon_0$) generated containing 
           $\tilde\chi^0_2$ in the final state, and the efficiencies 
           ($\epsilon_S$) for their detection using the two sets of
           optimized cutoffs as discussed in the text, for different 
           values of squark/gluino mass. The uncertainties are purely 
           statistical.}
  \label{tab:mc}
\end{table}

The detection efficiencies ($\epsilon_S$) for predictions from the 
supersymmetric models are given in Table~\ref{tab:mc},
along with the percentages ($\epsilon_0$) of generated events having 
$\tilde\chi^0_2$ in the final state. 
MC studies show that the overall efficiency varies by 4\% for different choices 
of $M_1$, $M_2$, $\tan\beta$, and $\mu$ that are consistent with
${\rm B}(\tilde\chi^0_2\rightarrow\gamma\tilde\chi^0_1)=1$
and $m_{\tilde\chi^0_2}-m_{\tilde\chi^0_1}>20$~GeV, the suggestions offered 
in Ref.\cite{kane}. Experimentally, the mass requirement is needed to ensure 
that photons from $\tilde\chi^0_2$ decays are reasonably energetic and can 
be detected with good efficiency. 
The total systematic error on the 
efficiency is 9\%, including uncertainties in photon identification 
efficiency (7\%), the choice of values of the supersymmetry parameters 
(4\%), and the jet energy scale (3\%).

We set a 95\% confidence level (C.L.) upper limit on
$\sigma\times {\rm B}\equiv\sigma(p\bar{p}\rightarrow\tilde q/\tilde g\rightarrow
\tilde\chi^0_2+X)\times {\rm B}(\tilde\chi^0_2\rightarrow\gamma\tilde\chi^0_1)$
using a Bayesian approach with a flat prior distribution for the signal cross 
section. The statistical and systematic uncertainties on the efficiency, 
the integrated luminosity, and the background estimate are included in 
the calculation of the limit, assuming Gaussian prior distributions. 
The resulting upper limit as a function of squark/gluino mass is tabulated in 
Table~\ref{tab:xslim}. 

\begin{table}[htbp]
  \squeezetable
  \begin{tabular}{c|cc|cc|cc}
   $m_{{\tilde q}/{\tilde g}}$ &\multicolumn{6}{c}{$\sigma\times{\rm B}$ (pb)} \\
                               &\multicolumn{2}{c|}{$m_{\tilde q}(=m_{\tilde g})$}
                               &\multicolumn{2}{c|}{$m_{\tilde g}(\ll m_{\tilde g})$}
                               &\multicolumn{2}{c} {$m_{\tilde q}(\ll m_{\tilde g})$} \\
   GeV & Theory & Limit & Theory & Limit & Theory & Limit \\ \hline
    150 & 83.4 & 2.0  & 24.1 & 2.6  & 8.51 & 1.8  \\
    200 & 12.1 & 1.1  & 3.48 & 1.6  & 1.59 & 0.9  \\
    250 & 2.37 & 0.57 & 0.51 & 0.63 & 0.43 & 0.58 \\
    300 & 0.53 & 0.39 & 0.12 & 0.44 & 0.12 & 0.38 \\
    350 & 0.13 & 0.37 & 0.02 & 0.37 & 0.03 & 0.32 \\
    400 & 0.04 & 0.36 & 0.01 & 0.37 & 0.01 & 0.32 \\
  \end{tabular} 
  \caption{The theoretical cross section $\sigma\times{\rm B}$ and our
           measured 95\% confidence level upper limit
           on $\sigma\times{\rm B}$ for different values of squark/gluino mass. 
           The predictions are calculated for $M_1=M_2=60$~GeV, $\tan\beta=2$,
           and $\mu=-40$~GeV.}
  \label{tab:xslim}
\end{table}

\begin{figure}[htbp]
  \epsfysize=3.0in\epsfbox{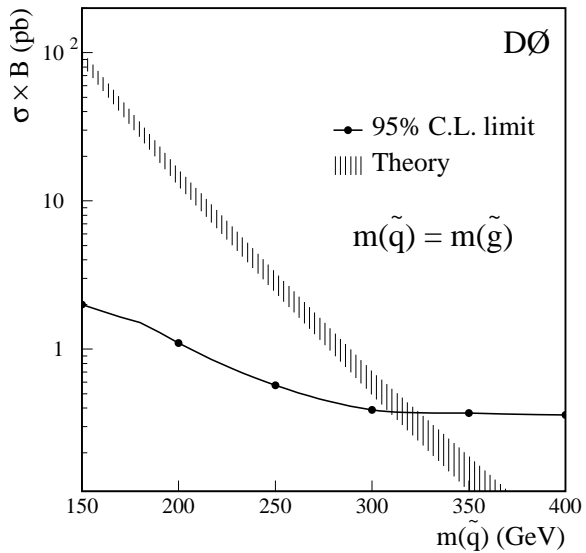}
  \caption{The 95\% C.L. upper limit on $\sigma\times{\rm B}$ as a function
           of $m_{{\tilde q}/{\tilde g}}$, assuming equal squark and gluino 
           masses. The hatched band represents the range of expected
           cross sections for different sets of MSSM parameters, 
           consistent with the constraints
           ${\rm B}(\tilde\chi^0_2\rightarrow\gamma\tilde\chi^0_1)=1$ and
           $m_{\tilde\chi^0_2}-m_{\tilde\chi^0_1}>20$~GeV. The inflection
           below 200~GeV in the limit curve is the intersection of the two
           curves using the two sets of optimized cutoffs discussed in the 
           text.}
  \label{fig:prl3}
\end{figure}

Figure~\ref{fig:prl3} shows the limit for the case where 
$m_{\tilde q}=m_{\tilde g}$, together with the leading-order theoretical cross 
section, calculated using the {\sc spythia} program with the CTEQ3L parton 
distribution functions\cite{cteq}. The renormalization scale ($\mu_{RS}$) is 
set to the average transverse energy ($\langle E_T \rangle $) of the outgoing 
partons in the calculation. The cross section varies by about $\pm 30\%$ if 
$\mu_{RS}=2\langle E_T \rangle $ or $\mu_{RS} = \langle E_T \rangle/2$ is used. 
The hatched band represents the range of predictions obtained by varying the 
supersymmetry parameters with the constraints that
${\rm B}(\tilde\chi^0_2\rightarrow\gamma\tilde\chi^0_1)=1$ and 
$m_{\tilde\chi^0_2}-m_{\tilde\chi^0_1}>20\ {\rm GeV}$, assuming
$\mu_{RS}=\langle E_T \rangle$. The intersection of the limit with the lower edge 
of the band is at $\sigma\times {\rm B}=0.38$~pb, leading to a lower limit 
for equal mass squarks and gluinos of 310~GeV at the 95\% C.L.

The effect of light sleptons on squark and gluino decays is studied 
by varying the slepton mass 
($m_{\tilde\ell}=m_{\tilde e}=m_{\tilde\mu}=m_{\tilde\tau}$) from 500~GeV to
80~GeV in the MC. For $m_{\tilde q}=m_{\tilde g}=300$~GeV MC events, the 
percentage $\epsilon_0$ increases by an additional 25\%.
Sleptons with mass below 80~GeV have already been excluded\cite{lep}.
The increase in $\tilde\chi^0_2$ production increases the mass limit by 
approximately 10~GeV. 

A light stop $\tilde t_1$ would also modify squark and gluino decays 
and would therefore affect $\tilde\chi^0_2$ production. 
If $m_{\tilde t_1}$ is lowered from 500~GeV to the lower experimental
limit of 80~GeV\cite{lep,d0stop}, a 15\% reduction in 
$\tilde\chi^0_2$ production is predicted.
This reduction lowers the limit for equal mass squarks and gluinos by 
about 6~GeV.

Following the above procedure, we obtain a lower limit for 
gluino (squark) mass of 240~GeV when squarks (gluinos) are heavy. 
Again, these limits vary by approximately 10~GeV if $\tilde t_1$ and/or 
sleptons are light. 


In summary, we have searched for an excess of $\gamma\rlap{\kern0.25em/}E_T$ 
events with two or more jets in $p\bar{p}$ collisions at $\sqrt{s}=1.8$~TeV.
Such events are predicted in the minimal supersymmetric standard model. 
We find that the number of observed \gmetjj\ events agrees well with that 
expected from background processes. Within the framework of the MSSM, with 
choices of parameters consistent with 
${\rm B}(\tilde\chi^0_2\rightarrow\gamma\tilde\chi^0_1)=1$ and 
$m_{\tilde\chi^0_2}-m_{\tilde\chi^0_1}>20$~GeV, we obtain a 95\% C.L. 
lower mass limit of 310 GeV for equal mass squarks and gluinos and of
240 GeV for squarks (gluinos) when gluinos (squarks) are heavy.
These limits constrain the models discussed in Ref.\cite{kane}, but do not 
exclude all of them.

We thank the staffs at Fermilab and collaborating institutions for their
contributions to this work, and acknowledge support from the
Department of Energy and National Science Foundation (U.S.A.),
Commissariat  \` a L'Energie Atomique (France),
Ministry for Science and Technology and Ministry for Atomic
   Energy (Russia),
CAPES and CNPq (Brazil),
Departments of Atomic Energy and Science and Education (India),
Colciencias (Colombia),
CONACyT (Mexico),
Ministry of Education and KOSEF (Korea),
and CONICET and UBACyT (Argentina).


\end{document}